# Formation enthalpy of $BaCe_{0.7}Nd_{0.2}In_{0.1}O_{2.85}$

M.Yu. Matskevich[1], N.I. Matskevich[1], Th. Wolf[2], A.N. Bryzgalova[1], T.I. Chupakhina[3], O.I. Anyfrieva[1], I.V. Vyazovkin[1]

[1]*Nikolaev Institute of Inorganic Chemistry, Siberian Branch of the Russian Academy of Science, Novosibirsk, 630090, Russia*

[2]*Karlsruhe Institute of Technology, Institute of Solid State Physics, Karlsruhe, D-76344, Germany*

[3]*Institute of Solid State Chemistry, Ural Branch of the Russian Academy of Science, Ekaterinburg, 620041, Russia*

## 1. Introduction

The compound $MCe_{1-x}R_xO_{3-\delta}$ (M = Ba and Sr, R = rare earth elements, x < 0.3) belongs to the perovskite-type oxides and shows a high protonic conductivity at elevated temperatures [1-11 ]. Proton conduction in doped barium or strontium cerates has been the subject of intense investigation in recent years because of the potential applications in fuel cell, sensor, and ceramic membrane technologies. In general, the transport properties of these materials depend critically on crystal structure and chemical stoichiometry. For perovskites the most important reaction leading to the formation of proton defects at moderate temperatures is the dissociative absorption of water, which requires the presence of ion vacancies. In the case of perovskite-type oxides the vacancies may be formed by substitution of up to 20% of the B-site cation by a lower-valent cation, i.e. $A(B_{1-x}R_x)O_{3-d}$ (R – rare earth element). Neodymium-doped barium cerates are believed to be the most conductive electrolytes. It is necessary to note that the dopant solubility limit is less than 20% of the available B sites, and therefore, the possibility of introducing protonic defects into the matrix is also limited. But it is important to increase the dopant solubility for further practical applications. All attempts to increase the dopant solubility limit were unsuccessful.

Another problem is stability of doped $BaCeO_3$ perovskites. The stabilities of these phases, particularly thermodynamic stability, are important in order to underpin further practical applications. The thermodynamic stability of complex oxides may have an impact on the mechanical stability of the microstructure of corresponding ceramics. Thermochemistry of this system involves at least three issues: (1) the stability of the mixed oxides with respect to their internal reactions within a material of fixed cation content, involving changes in oxygen stoichiometry, symmetry changes, and phase transitions; (2) stability with respect to other oxide phases with different cation stoichiometries in the multicomponent oxide system; (3) stability with respect to reactions with extraneous constituents such as water and carbon dioxide.

The investigation of proton conductivity in perovskite ceramics started more than 10 years ago. A number of reports were published with regard to the conduction of $BaCeO_3$-based oxides. However, there is still a lack of studies on other properties of these materials. Particularly, there are no thermodynamic values for compounds in the systems $BaO–CeO_2–R_2O_3–In_2O_3$ at all.

The paper is devoted to determination of formation enthalpy of $BaCe_{0.7}Nd_{0.2}In_{0.1}O_{3-\delta}$ phase. The series of other papers will be devoted to problem of thermodynamic stability in the system.

## 2. Experimental

We synthesized the $BaCe_{0.7}Nd_{0.2}In_{0.1}O_{3-\delta}$ sample for the first time. In the sample the field of solubility was increased by involving of indium.

Polycrystalline samples of $BaCe_{0.7}Nd_{0.2}In_{0.1}O_{3-\delta}$ were prepared by solid-state reaction. The synthesis was performed according to reaction: $BaCO_3 + 0.5x\ (In, R)_2O_3 + (1\text{-}x)CeO_2 = BaCe_{1-x}(In, R)_xO_{3-x/2} + CO_2$. Starting reagents were treated before synthesis at 600 ºC ($In_2O_3$), 800 ºC ($R_2O_3$) up to constant weight. A stoichiometric mixture of $BaCO_3$ (>99%, MERCK), $CeO_2$ (99.99%, Johnson Matthey GmbH, Alfa Products), $In_2O_3$ (82.1% In, Johnson Mattey, Materials Technology UK), $Nd_2O_3$ (99.9%, Reacton, Rare Earth Products, Division of Johnson Mattey Chemical LTD) was mixed in an agate mortar and ground for about 70 h with 10 intermediate regrounds in a Planetary Ball Mill (FRITSCH pulverisette 5. The rate was changed from 50 up to 200 tmp. Then the mixture was pressed (tablet ∅ 14 mm, press Herzog (5.5t)), placed in a furnace (Horizontal Tube Furnace, Item CTF 18/300, 1800 ºC) and heat treated in air according to the following regime: 800 ºC for 16 h, 1100 ºC for 10 h, 1400 ºC for 24 h, 800 ºC for 30 h. The phase purity was analyzed with X-ray diffractometer (STADI-P, Stoe diffractometer, Germany, Cu $K_{\alpha 1}$ or Mo $K_{\alpha 1}$ radiation). The samples were shown to be phase-pure ceramics with an orthorhombic structure (space group is Pmcn). The refined cell parameters obtained for $BaCe_{0.7}Nd_{0.2}In_{0.1}O_{3-\delta}$ are: $a$ = 0.8797 (9) nm, $b$ = 0.6203(6) nm and $c$ = 0.6205 (5) nm. The powder X-ray diffraction pattern of barium cerate doped by indium and neodymium oxides (experimental and differential) is presented in Figure 1.

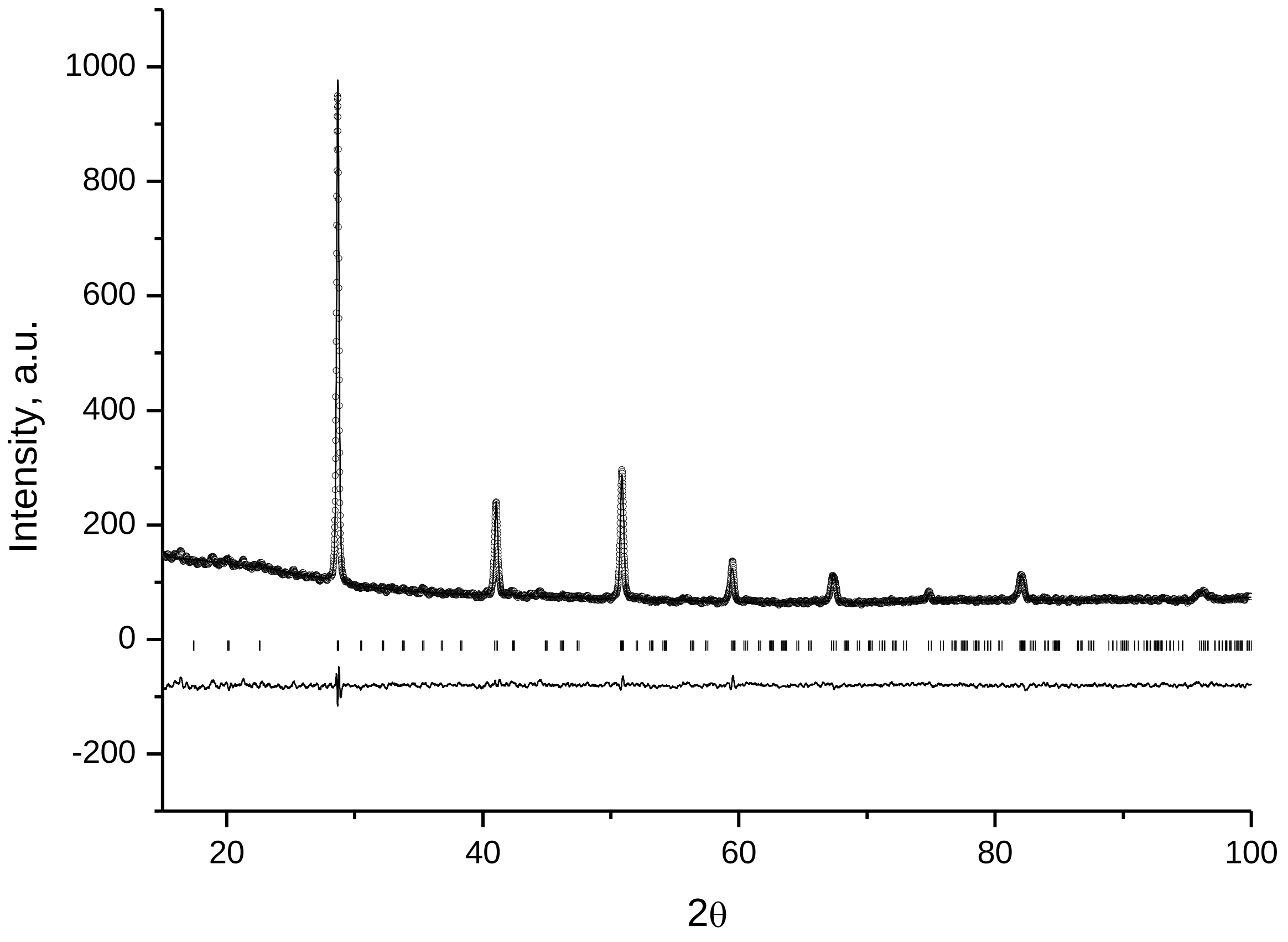

Figure 1. Powder X-ray diffractogram of $BaCe_{0.7}Nd_{0.2}In_{0.1}O_{3-\delta}$

Anhydrous $BaCl_2$ was prepared by drying $BaCl_2$ (CERAC, TM incorporated, USA, 99.9%) in argon at about 500 K. $CeCl_3$ was also purchased from CERAC (mass fraction is more then 0.999) and purified by vacuum sublimation in order to remove the lanthanide oxychloride impurities. For this purpose $CeCl_3$ was sublimated above the melting temperature (1143 K) in a vacuum better than $10^{-5}$ Pa. $NdCl_3$ was prepared from $Nd_2O_3$. $Nd_2O_3$ (99.99%, Purathem, STREM Chemicals, New buryport, USA) was dissolved in an excess of hydrochloric acid. Purified chlorine gas was bubbled through the solutions. Solution was then evaporated. Further drying was accomplished by evaporating under vacuum at about 350 K until the remaining chloride crystals appeared in composition. Final drying was accomplished by heating slowly in a hydrogen chloride atmosphere to a final temperature of 600 to 700 K. $InCl_3$ was synthesized from $Cl_2$ and In. Chlorine gas was passed over indium at temperature about 450 K. All manipulations with $CeCl_3$, $BaCl_2$, $NdCl_3$ and $InCl_3$ were performed in a dry box (pure Ar gas).

The solution calorimetric experiments were carried out in an automatic calorimeter with an isothermal jacket. The calorimeter consists of a Dewar vessel with a brass cover (V = 200 ml). The platinum resistance thermometer, calibration heater, cooler, mixer, and device to break the ampoules were mounted on the lid closing the Dewar vessel. The construction of the solution calorimeter and the experimental procedure are described elsewhere [12-13]. The resistance of platinum resistance thermometer was measured by high precision voltmeter Solartron 7061. The voltmeter was connected with the computer through interface and the program written in Matlab [14] at our laboratory. The program allows one to measure and record the temperature of the vessel, calibrate the instrument with precise injections of electrical energy and calculate calorimeter constants and enthalpies. The calorimetric vessel was maintained at 298.15 K with temperature drift less than 0.0003 °C for 10 h. Dissolution of potassium chloride in water was performed to calibrate the calorimeter. The obtained dissolution heat of KCl was 17.41 ± 0.08 kJ $mol^{-1}$ (the molality of the final solution was 0.028 mol $kg^{-1}$, T = 298.15 K). The literature data are: 17.42 ± 0.02 kJ $mol^{-1}$)[15], 17.47 ± 0.07 kJ $mol^{-1}$[16].

The amounts of substance used $BaCe_{0.7}Nd_{0.2}In_{0.1}O_{3-\delta}$ were about 0.08 g. All compounds were stored in a dry box to prevent interaction with moisture or $CO_2$.

In general, three techniques are used to determine the enthalpy of formation at 298.15 K of metal cerates: solution calorimetry, e.m.f. measurements and mass spectrometric Knudsen cell measurements [17-20]. In our paper we used solution calorimetry as an investigation method. 1 M HCl with 0.1 M KI was chosen as a solvent. KI was added to reduce $Ce^{+4}$ to $Ce^{+3}$.

The derivation of the enthalpy of formation of $BaCe_{0.7}Nd_{0.2}In_{0.1}O_{2.85}$ was done using the following scheme of thermochemical reactions (see, Table 1). The principal scheme is based on the dissolution of barium cerate doped by neodymium oxide and indium oxide as well as the mixture of barium chloride, cerium chloride, indium chloride and neodymium chloride in hydrochloric acid [HCl (sol)] with KI. The molar concentration of metal chlorides was the same as in paper [17]. A mixture of $BaCl_2$, $CeCl_3$, $NdCl_3$, $InCl_3$ was prepared in ratio 1: 0.7:0.2:0.1.

Table 1. Thermochemical cycle to determine the formation enthalpy of $BaCe_{0.7}In_{0.1}Nd_{0.2}O_{2.85}$

| Reaction | $\Delta_rH°_m$, кДж | Reference |
|---|---|---|
| 1. $BaCe_{0.7}In_{0.1}Nd_{0.2}O_{2.85}$(s) + (5.7HCl + 1.05KI)(sol) = ($BaCl_2$ + 0.7$CeCl_3$ + 0.1$InCl_3$ + 0.2$NdCl_3$ + 0.7KCl + 0.35$KI_3$ + 2.85$H_2O$)(sol) | –354.17 ±3.21 | This work |
| 3. $BaCl_2$(s) + 0.7$CeCl_3$(s) + 0.1$InCl_3$(s) + 0.2$NdCl_3$(s) = ($BaCl_2$ + 0.7$CeCl_3$ + 0.1$InCl_3$ + 0.2$NdCl_3$)(sol) | –146.85 ± 1.29 | This work |
| 3. 2.85$H_2$(g) + 1.425$O_2$(g) + (раствор 1) = 2.85$H_2O$(sol) | –814.65 ± 0.13 | [17] |
| 4. 1.05KI(s) + (раствор 1) = 1.05KI(sol) | +21.88 ± 0.46 | [17] |
| 5. 1.05K(s) + 0.525$I_2$(s) = 1.05KI(s) | –345.61 ± 0.18 | [17] |
| 6. 0.35K(s) + 0.525$I_2$(s) = 0.35$KI_3$(sol) | –105.96 ± 0.12 | [17] |
| 7. 0.7KCl(s) + (раствор 1) = 0.7KCl(sol) | +12.61 ± 0.04 | [17] |
| 8. 0.7K(s) + 0.35$Cl_2$(g) = 0.7KCl(s) | –305.53 ± 0.11 | [17] |
| 9. 2.85$H_2$(g) + 2.85$Cl_2$(g) + (раствор 1) = 5.7HCl(sol) | –936.85 ± 0.06 | [17] |
| 10. Ba(s) + $Cl_2$(g) = $BaCl_2$(s) | –855.15 ± 1.73 | [15] |
| 11. 0.7Ce(s) + 1.05$Cl_2$(g) = 0.7$CeCl_3$(s) | –742.38 ± 0.53 | [15] |
| 12. 0.1In(s) + 0.15$Cl_2$(g) = 0.1$InCl_3$(s) | –53.72 ± 0.84 | [15] |
| 13. 0.2Nd(s) + 0.3$Cl_2$(g) = 0.2$NdCl_3$(s) | –208.11 ± 1.26 | [15] |
| 14. Ba +0.1In + 0.2Nd + 0.7Ce + 1.425$O_2$ = $BaCe_{0.7}$ $In_{0.1}Nd_{0.2}O_{2.85}$ | –1607.99 ± 9.96 | This work |

The measured enthalpies of solution of $BaCe_{0.7}In_{0.1}Nd_{0.2}O_{2.85}$ and $BaCl_2$ + 0.7$CeCl_3$ + 0.1$InCl_3$ + 0.2$NdCl_3$ were determined as: $\Delta_{sol}H^o{}_1$(298.15 K) = –354.17 ±3.21 kJ/mol (n = 5), $\Delta_{sol}H^o{}_2$(298.15 K) = –146.85 ± 1.29 (n = 6). Errors were calculated for the 95% confidence interval using standard procedure of treatment of experimental data [21].

The measured enthalpies of dissolution were used for calculating the enthalpy of the reaction:

$Ba(s) + 0.1In(s) + 0.2Nd(s) + 0.7Ce(s) + 1.425O_2(g) = BaCe_{0.7}In_{0.1}Nd_{0.2}O_{2.85}$ (s)

according to the equation

$$\Delta_r H_{14}^o = -\Delta_{sol}H_1^o + \Delta_{sol}H_2^o + \Delta_{sol}H_3^o - \Delta_{sol}H_4^o - \Delta_{sol}H_5^o + \Delta_{sol}H_6^o + \Delta_{sol}H_7^o + \Delta_{sol}H_8^o - \Delta_{sol}H_9^o + \Delta_{sol}H_{10}^o + \Delta_{sol}H_{11}^o + \Delta_{sol}H_{12}^o + \Delta_{sol}H_{13}^o$$

Here, $\Delta_r H_{14}^o$ = $\Delta_f H^o$($BaCe_{0.7}In_{0.1}Nd_{0.2}O_{2.85}$, s, 298.15 K) = –1608.0 ± 10.0 kJ/mol is the standard formation enthalpy of barium cerate doped by neodymium and indium oxides. To calculate this value we used experimental data measured by us and literature data for formation enthalpies of different compounds and processes taken from Ref. [15, 17] and presented in Table 1.

Literature data for formation enthalpies of BaO, $CeO_2$, $Nd_2O_3$ and $In_2O_3$ taken from Ref. [15] were used to calculate the enthalpies of formation of $BaCe_{0.7}In_{0.1}Nd_{0.2}O_{2.85}$ from binary oxides as following:

$BaO(s) + 0.7CeO_2(s) + 0.1Nd_2O_3(s) + 0.05 In_2O_3(s) = BaCe_{0.7}Nd_{0.2}In_{0.1}O_{2.85}$ (s)

$\Delta_{ox}H^o$ (298.15 K) = –69.8 ± 10.1 kJ/mol

It is necessary to know the Gibbs energies (ΔG = ΔH - TΔS) to understand whether the $BaCe_{0.7}Nd_{0.2}In_{0.1}O_{2.85}$ phase is stable or unstable with respect to decomposition to BaO + $0.7CeO_2 + 0.1Nd_2O_3 + 0.05In_2O_3$ mixture. There are no entropies of $BaCe_{0.7}Nd_{0.2}In_{0.1}O_{2.85}$ phase in literature. These values were estimated using entropies of $Nd_2O_3$, $BaCeO_3$, $CeO_2$, $In_2O_3$ taken from references [15]. The Gibbs free energy and the formation enthalpy from binary oxides are practically the same.

So, above-mentioned complex oxide ($BaCe_{0.7}Nd_{0.2}In_{0.1}O_{2.85}$) is thermodynamically stable with respect to their decomposition into binary oxides at room temperatures. It is not obvious result for this class of compounds because there is a discussion about thermodynamic stability of $BaCeO_3$[22-24]. There is a paper which claims that $BaCeO_3$ is instable with respect to its decomposition into BaO and $CeO_2$ [22].

In our case we have values at the level –70 kJ/mol which are higher than the usual values for complex oxides but lower than formation enthalpies from binary oxides for HTSC materials ($\Delta_{ox}H$ for $RBa_2Cu_3O_{7-x}$ phase (R – rare earth element) is about –200 kJ/mol). It is possible to compare $\Delta_{ox}H^o$ for barium cerate doped by neodymium and indium oxides ($\Delta_{ox}H^o$ ($BaCe_{0.7}Nd_{0.2}In_{0.1}O_{2.85}$, 298.15 K) = –69.8 ± 10.1 kJ/mol) with $\Delta_{ox}H^o$ for barium cerate doped by only neodymium oxide ($\Delta_{ox}H^o$ ($BaCe_{0.8}Nd_{0.2}O_{2.9}$, 298.15 K) = –30.4 ± 4.8 kJ/mol). The data for $BaCe_{0.8}Nd_{0.2}O_{2.9}$ were taken from our earlier paper [11]. It is possible to see that the stability is increasing with adding of indium oxide in sample.

## Conclusions

In this paper for the first time we synthesized the compound $BaCe_{0.7}Nd_{0.2}In_{0.1}O_{2.85}$ by solid-state reaction. The phase has orthorhombic structure (space group Pmcn). We also measured the standard formation enthalpies of $BaCe_{0.7}Nd_{0.2}In_{0.1}O_{2.85}$ by solution calorimetry in 1 M HCl with 0.1 M KI. We determined the stability of Nd(In)-doped barium cerate with respect to mixtures of binary oxides. On the basis of these data we established that above-mentioned mixed oxide is thermodynamically stable with respect to their decomposition into binary oxides at room temperatures. We also established that $BaCe_{0.7}Nd_{0.2}In_{0.1}O_{2.85}$ oxide is thermodynamically favoured than $BaCe_{0.8}Nd_{0.2}O_{2.9}$.

## Acknowledgments

This work is supported by Karlsruhe Institute of Technology (Germany), Russian Fund of Basic Research (Project № 12-08-31556 for young scientists) and Program of Fundamental Investigation of Siberian Branch of the Russian Academy of Sciences.